\documentclass[runningheads]{llncs}
\usepackage[T1]{fontenc}
\usepackage{graphicx}
\usepackage{hyperref}
\usepackage{color}
\graphicspath{ {./images/} }

\begin{document}
\title{In Situ In Transit Hybrid Analysis with Catalyst-ADIOS2}
\author{François Mazen\and
Louis Gombert\and
Lucas Givord\and
Charles Gueunet
\authorrunning{F. Mazen et al.}
\institute{Kitware Europe\\
\email{\{name.surname\}@kitware.com}}}
\maketitle              %
\begin{abstract}
In this short paper, we present an innovative approach to limit the required bandwidth when transferring data during in transit analysis. This approach is called \emph{hybrid} because it combines existing in~situ and in~transit solutions. It leverages the stable ABI of Catalyst version~2 and the Catalyst-ADIOS2 implementation to seamlessly switch from in~situ, in~transit and hybrid analysis without modifying the numerical simulation code. The typical use case is to perform data reduction in situ then generate a visualization in transit on the reduced data. This approach makes the numerical simulation workflows very flexible depending on the size of the data, the available computing resources or the analysis type. Our experiment with this hybrid approach, reducing data before sending it, demonstrated large cost reductions for some visualization pipelines compared to in situ and in transit solutions. The implementation is available under an open source permissive license to be usable broadly in any scientific community.

\end{abstract}
\section{Context}

In situ analysis is a technique to perform data exploration and visualization during the execution of a numerical simulation, in order to reduce I/O bottleneck~\cite{ref_in_situ_method}. This technique is de facto standard when a simulation operates with large data and during several time steps. In transit analysis is an extension of in situ analysis by moving the data to a dedicated node for analysis. This circumvents many drawbacks of in situ analysis like reducing the simulation halt time, and performing analysis on dedicated visualization nodes~\cite{ref_catalyst-adios2}. These visualization nodes could use dedicated hardware like powerful GPUs which are not necessarily available on the nodes running the numerical simulation. However, in transit analysis requires moving the dataset over the network to reach the end point where the analysis will be performed. Thus, in transit technique is not scalable, especially when reaching exascale. In this case, the simulation output data are so large that it is nearly impossible to transfer the data in an acceptable amount of time. Thus, the data transfer becomes the bottleneck.

Many numerical simulations mitigate the data transfer problem by computing data reduction at the end of a step. For example, OpenFOAM could generate specific data reduction~\cite{ref_openfoam-user-guide}. Some simulation codes could embed visualization routines like in DualSPHysics~\cite{ref_dualsphysics} which leverage the Visualization ToolKit (VTK) to perform isosurface extraction on-the-fly. However these solutions lack flexibility because the scientists are limited by the available features made available through simulation codes.

Kress \& al.~\cite{cloverleaf} identified several situations where the Visualization Cost Efficiency Factor (VCEF) is not sufficient to achieve cost savings. In particular, the case where the transfer cost is bigger than executing the visualization in-line.

\section{In Situ and In Transit Hybrid Analysis}

In order to benefit from in transit analysis while reducing the cost of data transfer, we developed a combination of in situ and in transit technologies. Catalyst version 2 is a recent standard to instrument numerical simulation code in a portable way thanks to its stable Application Binary Interface (ABI)~\cite{ref_catalyst2}. As Catalyst is just a protocol to exchange data, there are several implementations or backends available. Catalyst can also work with externally managed data pointers, which allow sharing data between the simulation and the visualization software without performing any copy. The reference implementation is Catalyst-ParaView, where the data are internally transformed to VTK data objects and processed through a ParaView python analysis pipeline. Another Catalyst implementation is Catalyst-ADIOS2~\cite{ref_catalyst-adios2} which uses the efficient data movement capabilities of ADIOS2~\cite{ref_adios2} to perform in transit analysis. Catalyst uses Conduit Blueprint as the backbone of data description across the various backends. 

Our hybrid approach combines both of these Catalyst implementations. The data are first reduced in situ thanks to the Catalyst-ParaView implementation on the simulation nodes, then the reduced data are passed to the Catalyst-ADIOS2 implementation to be moved and then replayed with the Catalyst-ADIOS2 Replay mechanism on the visualization nodes. This replay executable then passes the reduced data to a Catalyst-ParaView implementation inside of the Visualization Cluster to perform the final analysis with dedicated resources, for example using GPUs. Unlike classic in~situ, this last analysis part on the visualization cluster does not block the running simulation on the simulation cluster.

\begin{figure}
\includegraphics[width=\textwidth]{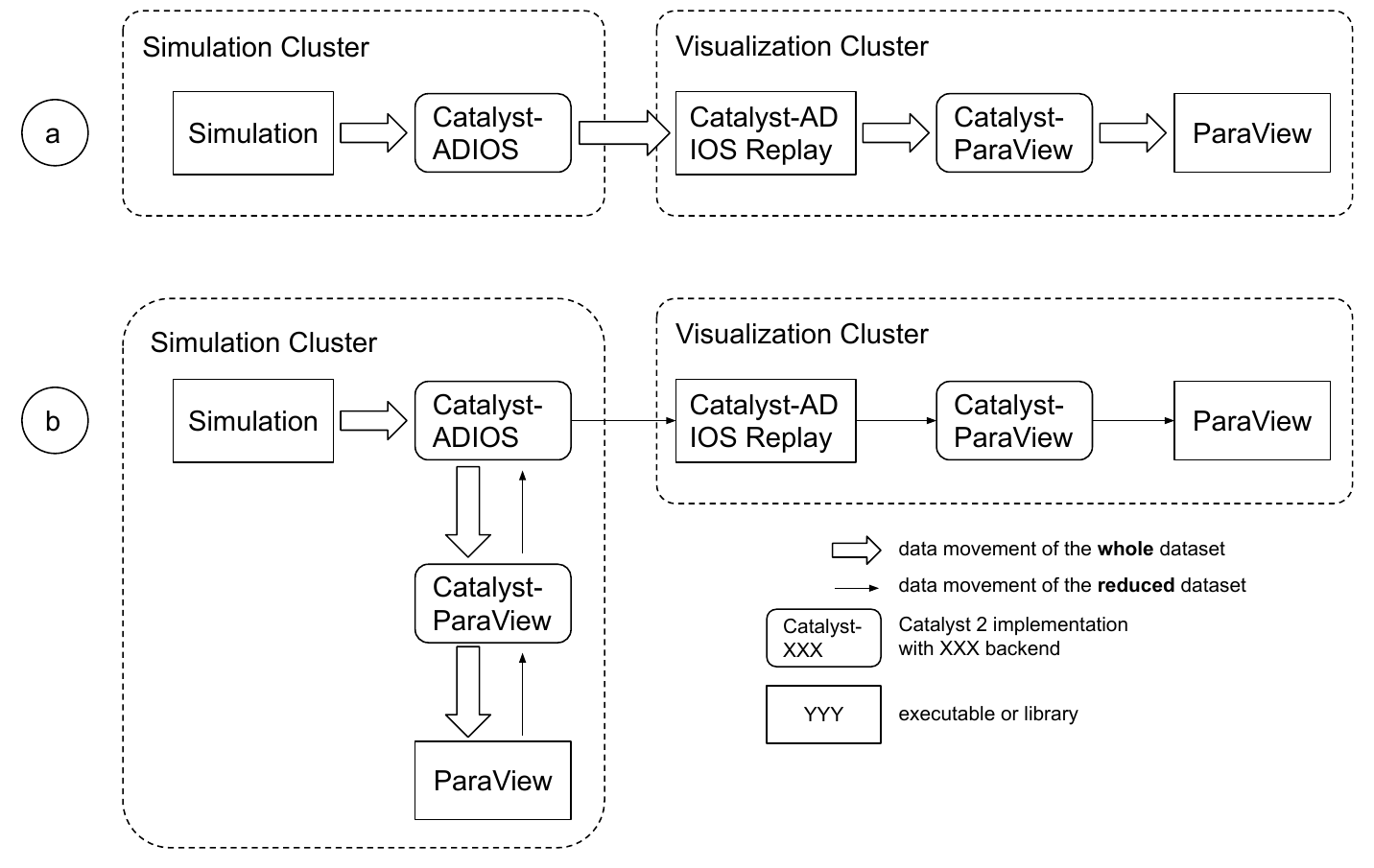}
\caption{Comparison of data flow between in transit solution with Catalyst-ADIOS2 (a) and our new hybrid in situ in transit approach (b), where only reduced data are sent over the network.} \label{fig-data-flow}
\end{figure}

In this configuration, the numerical simulation code does not have to change because the entry point is still the Catalyst-ADIOS2 implementation on the simulation side. Our work was to upgrade the Catalyst-ADIOS2 project to orchestrate the combination of data reduction then data movement, as described in figure~\ref{fig-data-flow}.

\begin{figure}
	\includegraphics[width=\textwidth]{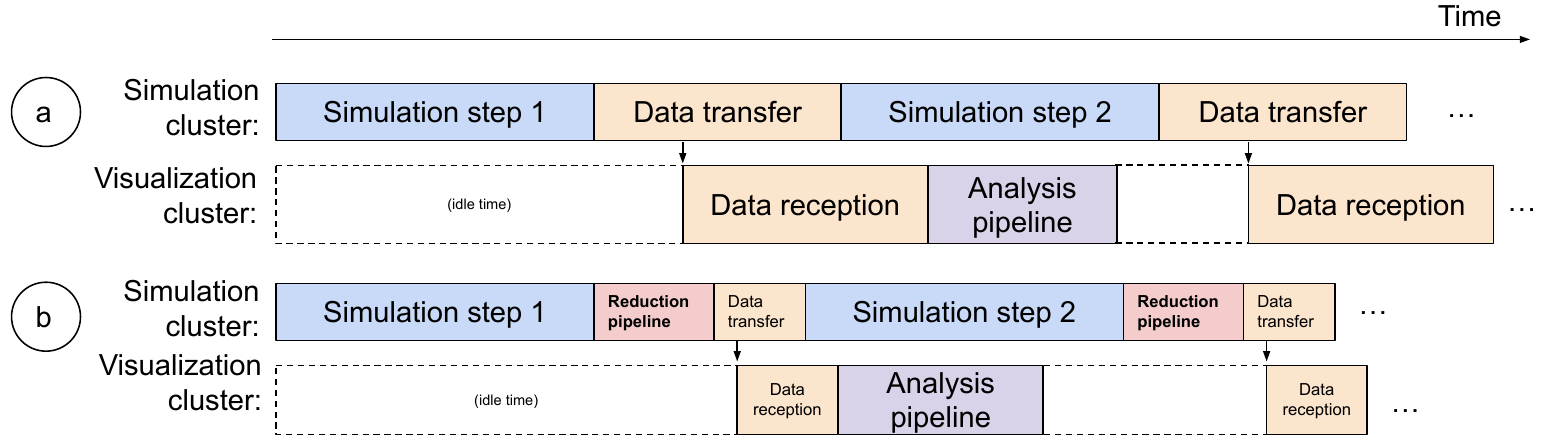}
	\caption{Schematic timeline comparison of in transit analysis (a) and our new hybrid in situ in transit approach (b). Performance gain is expected when data reduction time and reduced data transfer time (b) is less than full data transfer time (a).} \label{fig-timeline}
\end{figure}

This approach compels the numerical simulation user to prepare two pipelines to be executed by the Catalyst-ParaView libraries. The first pipeline is for the data reduction in situ and the second one is for the final analysis on the reduced data on the simulation cluster. Thanks to the stable ABI of Catalyst version 2, the numerical simulation code does not depend on the chosen Catalyst backend. The Catalyst-ParaView and ParaView versions could have different configurations and be optimized for their running hardware. In a typical situation where the simulation cluster contains only CPUs, the ParaView instance used by the numerical simulation nodes would have to use all the available CPU-cores power with SMP backend like TBB or OpenMP, with headless and/or offscreen rendering using EGL or OSMesa. On the other hand, the ParaView instance on the Visualization cluster would leverage the GPU capabilities with dedicated rendering backends like NVidia IndeX~\cite{ref_nvidia_index}.

In practice, we had to face two technical challenges. The first one was to load the Catalyst-ParaView implementation inside the Catalyst-ADIOS2 library to perform the data reduction using ParaView algorithms. The goal was to let the simulation use the same implementation of Catalyst, such as we did for the in transit case introduced in~\cite{ref_catalyst-adios2}. The second challenge was to get the output of the ParaView data reduction pipeline back to the Catalyst-ADIOS2 library. For this, we had to improve the existing \emph{simulation steering} mechanism of Catalyst-ParaView by creating a new Steering Extractor filter. The output of this extractor is  serialized as a conduit node and sent back to Catalyst-ADIOS2 which ordered the execution of the reduction pipeline. Then, the reduced data is passed to ADIOS2 to be sent over the network to the visualization cluster. The process is entirely handled by the Catalyst-ADIOS2 library which requires very little code modification of the simulation.

The final implementation is publicly available under a permissive open source license at the Kitware's public Gitlab instance~\cite{ref_adios_catalyst_code}.

\section{Use Case and Benchmark}

It has been demonstrated before~\cite{cloverleaf} that some simulation workflows benefit from in~transit compared to in~line visualization. As depicted in figure~\ref{fig-timeline}, in situations where the visualization pipeline takes less time to execute than a simulation step, in transit saves time compared to in line. In the latter case, the time taken by the data transfer from the simulation cluster to the visualization cluster is shorter than the time it would take to perform the visualization in line. This is usually the case for visualization pipelines that involve computation-heavy algorithms such as isocontouring or volume rendering. However, cost savings are harder to achieve with larger data where the transfer time is important. Our approach is aimed to help saving costs in these situations by performing a data reduction in the simulation cluster before sending the data. Typically, this reduction can be a simple slice on the simulation data, or another type of dimension reduction. This in line reduction does not requires full simulation data movement, leveraging Catalyst's ability to manipulate pointers to data managed externally by the simulation.

For our benchmark, we considered that the visualization pipeline has two separable steps: data reduction (size, dimensionality, clustering, number of fields) and visualization of the reduced data. Compared to in transit analysis, we can achieve cost savings if the in line data reduction takes less time than sending the whole data through ADIOS2. All in all, the hybrid workflow is interesting to save costs for simulations that output data large enough to notice the synchronous transfer time, which takes resources on both clusters, and where the visualization is complex and heavy enough to justify using separate nodes for visualization.

\begin{figure}
  \includegraphics[width=\textwidth]{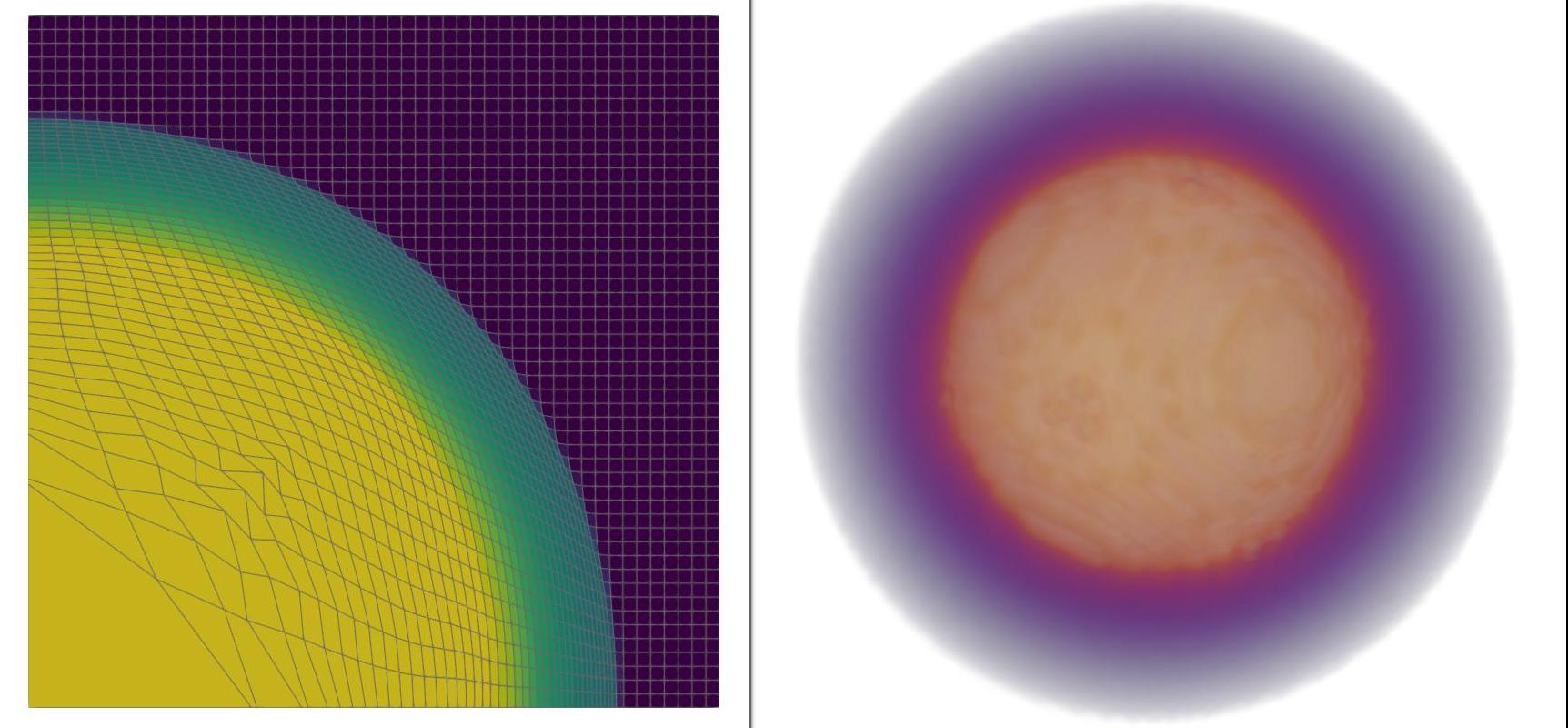}
  \caption{Results of the `slice' (left) and `resampled volume' (right) hybrid rendering pipelines for timestep 1662 for a reduced $50^3$ cell simulation mesh}\label{fig-renders}
\end{figure}

To evaluate our solution, we used the Livermore Unstructured Lagrangian Explicit Shock Hydrodynamics (LULESH) simulation code~\cite{ref_lulesh} which solves a simple Sedov blast problem, and written to run on multi-node cluster. We introduce two different pipelines: one performing a slice on the 3D data showing the wavefront of the blast, and another one transforming the unstructured mesh to a smaller one with a regular structure, on which we perform volume rendering to show the shock wave propagation. The result of those is shown figure~\ref{fig-renders}.

Our testbed involves a single machine using a 16 core-12th Gen Intel i9 CPU and 128GB of RAM. We run the LULESH simulation over several dozen time steps with a data size of $220^3$, around 10 million cells. The \emph{slice} reduction pipeline reduces the data transferred between the simulation and visualization cluster from $220^3$ cells to $220^2$, while the \emph{resampling} one transfers a $30^3$ structure for volume rendering. These reduction pipelines also filter out the variable fields that will not be used for visualization in order to decrease the amount of data transferred. On our test machine, we allocated 8 processes to run the simulation and another 8 to run the visualization. Due to current technical limitation in Catalyst-ADIOS2 with empty MPI rank, the \emph{slice} pipeline was run on a single core.

\begin{figure}
  \includegraphics[width=\textwidth]{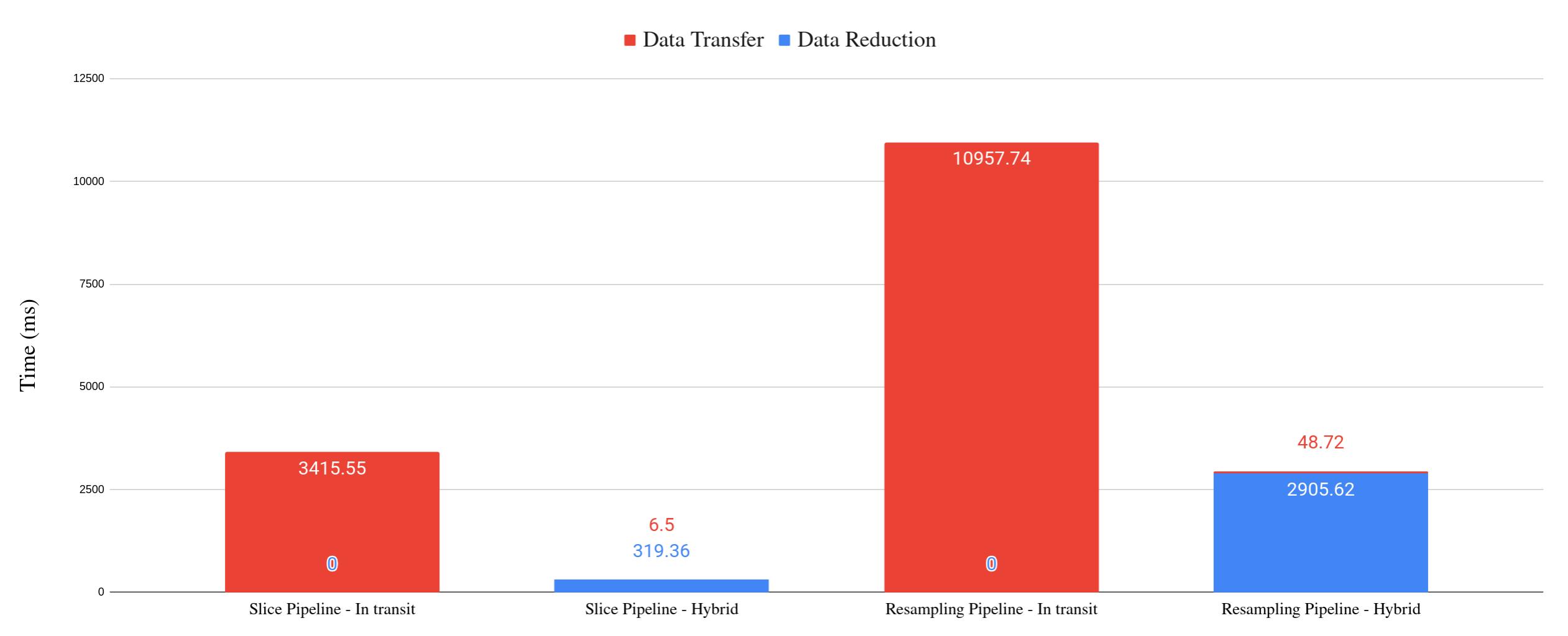}
  \caption{Comparison of the time taken by data reduction and data transfer for both pipelines on the simulation cluster in average by timestep, for the hybrid and the in transit cases.}\label{fig-benchmark}
\end{figure}

In order to verify our hypothesis that our hybrid approach saves time overall over in transit analysis, we measured the time taken by each MPI rank at each timestep to process the simulation, perform the reduction and send the data using ADIOS2. We compared our hybrid pipeline to the equivalent pipeline performed in transit. The results presented figure~\ref{fig-benchmark} and table~\ref{tab-benchmark} show the time taken by each pipeline for each of these steps averaged for all timesteps. In the case of the \emph{slice} pipeline, the data transfer time becomes negligible in the hybrid case, and the data reduction takes more than ten times less time than the full data transfer in the in transit case. For the volume rendering pipeline involving a data size reduction using resampling, the reduction is 3.7 times faster than full dataset transfer. In both cases, our approach saved computation time (16\% and 22\%) on the simulation cluster compared to classic in transit analysis.

\begin{table}
	\caption{Averaged compute time of time steps in milliseconds for different parts of the in transit and hybrid analysis.}\label{tab-benchmark}
	\centering
	\begin{tabular}{|l|l|l|l|l|}
		\hline
		&  \multicolumn{2}{c|}{\emph{Slice} Pipeline} & \multicolumn{2}{c|}{\emph{Resampling} Pipeline} \\

		\cline{2-5} & In Transit & Hybrid & In Transit & Hybrid \\
		\hline
		Simulation Time (ms)	& 	15860 	&	15835 &	33539 & 	31677\\
		\hline
		Reduction Time (ms)		&	0 		&	319 &	0 & 	2905\\
		\hline
		Data Transfer Time (ms)&	3415 &		6.56 &	10957 & 48.7 \\
		\hline
		\hline
		Total Time (ms)&			19275	& 	16161 &	44497 & 	34632 \\
		\hline
		Total Gain & \multicolumn{2}{c|}{16.16\%} & \multicolumn{2}{c|}{22.17\%}\\
		\hline
	\end{tabular}
\end{table}

For reproducibility purpose, the code used for the experiment is available under a permissive open~source license at~\cite{ref_adios_catalyst_hybrid_experiment}.

\section{Limitations and Perspectives}

Beside the experimentation which validated our approach to improve the performance of large simulations, we encountered several limitations.

First, the scientists have to prepare two pipelines, one for the data reduction and one for the visualization. In~situ analysis is notabily known for forcing the user to anticipate the shape of the generated data before the simulation, which is sometime impossible. Some techniques have been proposed to circumvent this issue, for example by optimizing the camera placement based on the entropy of the dataset~\cite{ref_optimizing-camera-placement}.

In our experiment, we reduced the size of the data via resampling to image which reduced the accuracy of the result. Hence, this approach should be evaluated and adapted to the expected analysis results. For example, if the goal is to check the general shape of the result then aggresive dimension reduction would be a good strategy. However for high accuracy results, the data reduction should be choosen and tuned carrefully to not introduce any quality loss.

Another limitation is that the workflow adds a new processing block, the Catalyst-ParaView in~situ processing, which limits the range of supported dataset and field types. In practice, only scalar fields and explicit topologies are supported by the framework so far, due to the current capabilities of Catalyst-ADIOS2 and Catalyst-ParaView dataset serialization.

The communication between the modules are done via the Catalyst API. Thanks to the strict ABI compatibility, the user can tune the data reduction library, currently the ParaView library, for the simulation cluster. For example, it could use an OSMesa implementation of ParaView in order to run on CPU nodes only, without forcing to use costly GPUs. As a perspective, we could envision to replace this part by some AI enhanced routines to speed-up the data reduction, without changing other processing blocks in the workflow.

Similarly, the end points could be any processing type. In our benchmark, we used the classic picture generation method to validate the correctness of the results, but we could imagine to perform advanced processing like super-resolutions of the data,  Machine Learning training with the reduced dataset or specific rendering techniques which require dedicated hardware like powerful GPUs or FPGAs.

Another interesting perspective is to use the data reduction to extract region of interest in the dataset, or compute data clustering for urgent decision making~\cite{ref_vestec}. The flexibility of the python pipeline of Catalyst would allow advanced detection of relevant parts to send in~transit. A typical use case have been described in the Inshimtu introduction~\cite{ref_inshimtu_blog} where the region of interest of the Chapala cyclone is dynamicaly extracted in situ.

\section{Conclusion}

We demonstrated that the main performance drawback of in transit analysis at scale, which is the data transfer time, could be mitigated with our hybrid approach. The data is first reduced in line, then the reduced data is sent over the network to the rendering nodes. Simple data reduction like resampling to a smaller dataset or discarding unused field shows very large performance gain. The innovative key point of this approach is to leverage the stable ABI of Catalyst version 2. It means that the simulation code would likely not be modified to benefit the hybrid approach, and that the flexible python pipeline does not limit the type of data reduction to perform. Finally, our approach demonstrates that the in transit analysis could be scalable by controlling the amount of transfered data in a flexible way. The proposed solution is freely available under a permissive open source license to foster reproducibility and broad adoption~\cite{ref_adios_catalyst_code}.

\end{document}